\documentclass[twocolumn,prl,aps,superscriptaddress]{revtex4-1}
\usepackage{epsf}
\newcommand{\unit}[1]{\ensuremath{\,\mathrm{#1}}} 
\usepackage{gensymb,longtable}
\usepackage{graphicx,color}

\usepackage{amssymb}
\usepackage{amsmath}
\usepackage{graphicx}
\usepackage{placeins}
\usepackage{booktabs}
\usepackage{array}
\usepackage[pdftex]{hyperref} 
\usepackage{amsthm}
\usepackage{float}
\usepackage{color,longtable}															 
\usepackage{colortbl}
\definecolor{hellgrau}{rgb}{0.95,0.95,0.95}
\hypersetup{colorlinks=true, linkcolor=black, citecolor=black, filecolor=black, urlcolor=black, pdftitle={Proposal}}
\usepackage{gensymb}

\usepackage{geometry}
\begin{document}

\author{S. Tesoro}
\affiliation{Cavendish Laboratory, University of Cambridge, JJ Thomson Avenue, CB3 0HE Cambridge, UK}
\author{K. G\"opfrich}
\affiliation{Cavendish Laboratory, University of Cambridge, JJ Thomson Avenue, CB3 0HE Cambridge, UK}
\author{T. Kartanas}
\affiliation{Cavendish Laboratory, University of Cambridge, JJ Thomson Avenue, CB3 0HE Cambridge, UK}
\author{U. F. Keyser}
\affiliation{Cavendish Laboratory, University of Cambridge, JJ Thomson Avenue, CB3 0HE Cambridge, UK}
\author{S. E. Ahnert}
\affiliation{Cavendish Laboratory, University of Cambridge, JJ Thomson Avenue, CB3 0HE Cambridge, UK}

\title{Non-deterministic self-assembly with asymmetric interactions} 

\begin{abstract}
We investigate general properties of non-deterministic self-assembly with asymmetric interactions, using a computational model and DNA tile assembly experiments. By contrasting symmetric and asymmetric interactions we show that the latter can lead to self-limiting cluster growth. Furthermore, by adjusting the relative abundance of self-assembly particles in a two-particle mixture, we are able to tune the final sizes of these clusters. We show that this is a fundamental property of asymmetric interactions, which has potential applications in bioengineering, and provides new insights into the study of diseases caused by protein aggregation.
\end{abstract}

\maketitle

\section{Introduction}

Self-assembling systems are ubiquitous in nature, with many examples in biology, chemistry, and physics, including protein complexes \cite{Levy:2006ez,Perica:2012jz,Ahnert:2015eb}, DNA tiles \cite{winfree1998design,Wei2012d}, colloids \cite{Rossi:2011gl}, micelles \cite{israelachvili1994self}, and diffusion-limited aggregation (DLA) \cite{Witten:1983wm}. While most chemical and physical self-assembly processes tend to be non-deterministic, most self-assembly processes in biology are deterministic, as the development and functioning of biological organisms requires that biological structures, from the molecular to the macroscopic, are formed repeatedly and accurately. The most prominent example is the enormous variety of protein complexes \cite{Levy:2006ez,Perica:2012jz,Ahnert:2015eb}, which fulfil important biological functions in all species, from bacteria to humans. For protein complexes to function in their respective context, their physical structures have to be correct, which means that the assembly process has to be deterministic in the sense that it must always lead to the same final structure. Mutations of the genome can result in protein misfolding, which in turn can lead to protein complex mis-assembly and uncontrolled, non-deterministic protein aggregation. This is the hallmark of a number of diseases, such as Alzheimer's and Parkinson's \cite{Selkoe:2004uc} and sickle cell anemia \cite{Bunn:1997de}. Non-deterministic self-assembly phenomena are therefore biologically relevant, but most models in the literature tend to focus on the specific context of particular proteins that aggregate. 

Here we study the general properties of non-deterministic assembly using both a computational model and DNA self-assembly experiments. In particular, we focus on asymmetric interactions between the self-assembling particles. Asymmetric interactions consist of interfaces with two different interacting surfaces, and are commonplace in biology. Contrasting these with the symmetric interactions, which are interfaces that self-interact and are found in many non-biological aggregation phenomena \cite{Rossi:2011gl,israelachvili1994self,Witten:1983wm}, we show that asymmetric interactions can result in self-limiting growth. This phenomenon is largely independent of the system size and strongly relies on the local steric properties of the asymmetric cluster aggregation process, which we can observe in our simulations and DNA tile experiments. Furthermore, we show that the relative proportions of two different self-assembling particles can allow us to tune the size of the aggregates.  

\section{Computational model}

\begin{figure}[h]
\includegraphics[width=0.45\textwidth]{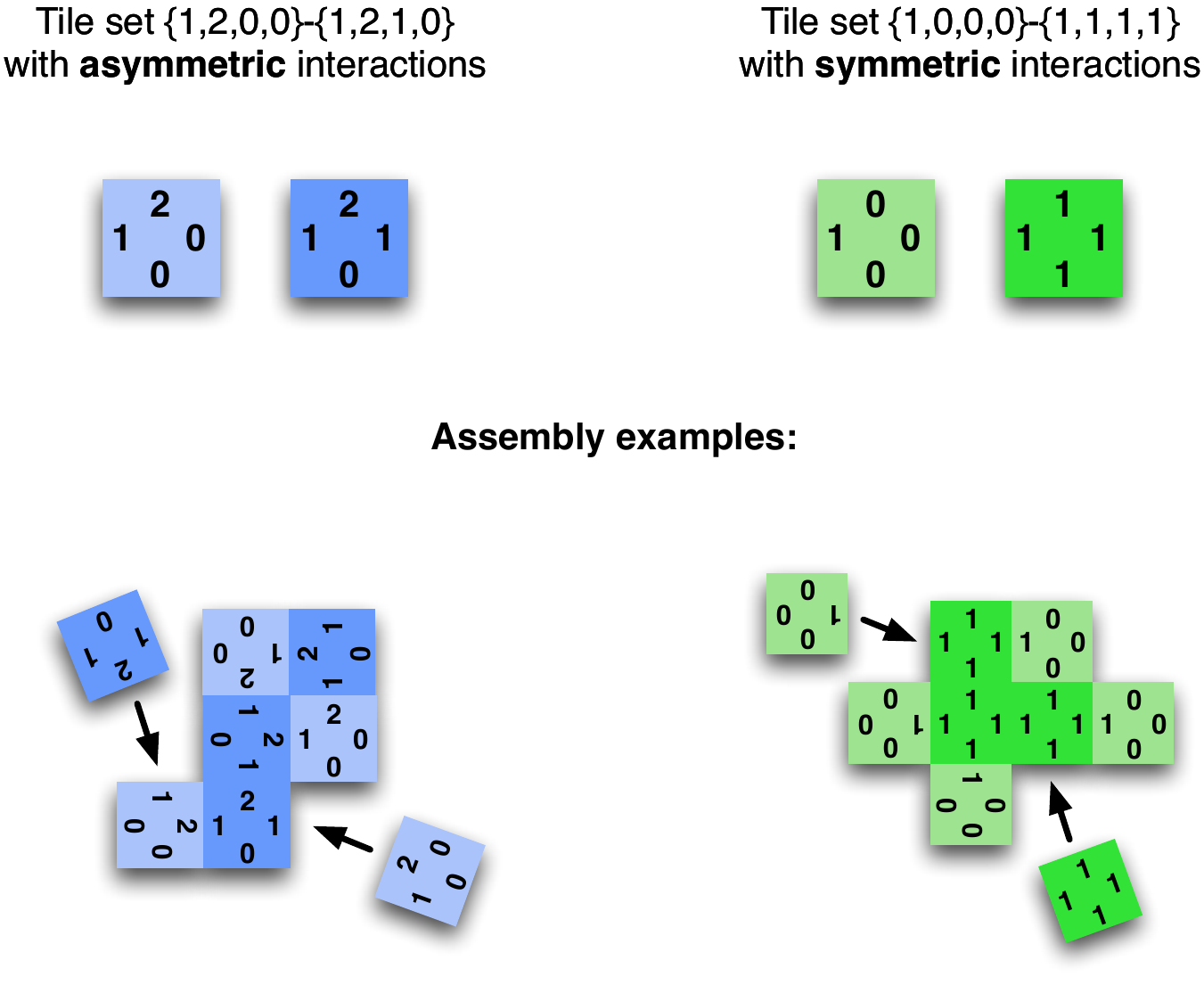}
\caption{Illustration of the two tile sets $\{1,2,0,0\}$-$\{1,2,1,0\}$ and $\{1,0,0,0\}$-$\{1,1,1,1\}$, which are studied under single- and multi-seed growth conditions. Tile set $\{1,2,0,0\}$-$\{1,2,1,0\}$ (left, blue) has asymmetric interactions, while tile set $\{1,0,0,0\}$-$\{1,1,1,1\}$ (right, green) has symmetric ones.}\label{tilesets}
\end{figure}

Our computational model is based on a two-dimensional lattice model of self-assembly introduced and studied in previous work \cite{Modularity,Evo,green,ARXIV2015}. In the model we have square tiles of two types, $A$ and $B$, which have interactions (termed 'colours') on their four faces, represented by integer numbers 0, 1 and 2. 0-faces are neutral (meaning they do not interact), while 1- and 2-faces can interact with each other in two ways. Under symmetric interactions, each colour is attracted to itself, i.e. 1 to 1, and 2 to 2. Under asymmetric interactions, 1 and 2 are attracted to each other, but not to themselves. The interactions are short-ranged and infinite in strength. We will study two tile sets of two tiles each: $\{1,0,0,0\}$-$\{1,1,1,1\}$ and $\{1,2,0,0\}$-$\{1,2,1,0\}$, where each bracket contains the colours, denoted in clockwise order, of the four faces on each of the tiles, and the hyphen connects the two tiles in a set. The former tile set will be studied under symmetric, and the latter under asymmetric interactions. Both are depicted in Figure \ref{tilesets}. Tile sets such as these have been studied in detail in previous work \cite{Modularity,Evo,green,ARXIV2015} with a single seed tile as the starting point. In the single-seed model, growth is initiated with a single seed tile, and further tiles are added one by one, randomly attaching to free interactive faces on the perimeter of the growing cluster. In sets of two or more tiles the relative concentrations of the tiles can greatly influence the types of structures formed. This is the case if the tile sets are non-deterministic, meaning that the tiles can attach to each other in more than one configuration \cite{Evo,ARXIV2015}.  
With single-seed initial conditions both $\{1,0,0,0\}$-$\{1,1,1,1\}$ and $\{1,2,0,0\}$-$\{1,2,1,0\}$ tile sets show a transition from bounded to unbounded growth, dependent on the relative concentrations of the tiles \cite{ARXIV2015}.

However, while single-seed initial conditions can be interesting in a deterministic context \cite{Modularity,Evo,green}, they are less realistic in biological and experimental conditions, where multi-seed growth conditions predominate. For this reason, we here introduce a model that allows multiple seed tiles as starting points. We show that for the tile set $\{1,2,0,0\}$-$\{1,2,1,0\}$ with asymmetric interactions, this initial condition does not result in a transition from bounded to unbounded growth, but instead in self-limiting cluster growth. Furthermore, the relative proportions of the two tiles can be used to tune the size of the final clusters. We confirm these results of our computational model by reproducing these two tiles in the form of self-assembling DNA tiles, for which we observe the same self-limiting growth and tunable cluster size in the case of asymmetric interactions.
For the tile set $\{1,0,0,0\}$-$\{1,1,1,1\}$ with symmetric interactions both the computational model and the DNA tile experiments result in a transition from bounded to unbounded growth as a function of the relative tile populations. In the computational model this behaviour occurs in both single-seed and multi-seed initial conditions, and has already been discussed in some detail in \cite{ARXIV2015}.

We can describe the growth behaviours of two-tile sets in terms of the relative proportions of the two tiles. For example, if $f$ is the proportion of the second tile and $1-f$ the proportion of the first tile in a two-tile set, we can define $f_c$, the critical value of $f$ at which growth goes from bounded to unbounded in the single-seed model, as published in \cite{ARXIV2015}. For the tile sets studied here these values are $f_c=0.53$ for the tile set $\{1,2,0,0\}$-$\{1,2,1,0\}$ with asymmetric interactions, and $f_c=0.25$ for the tile set $\{1,0,0,0\}$-$\{1,1,1,1\}$ with symmetric interactions. Much more detail about these results and the single-seed model can be found in \cite{ARXIV2015}.

\begin{figure}[h!]
  \centering
    \includegraphics[width=0.4\textwidth]{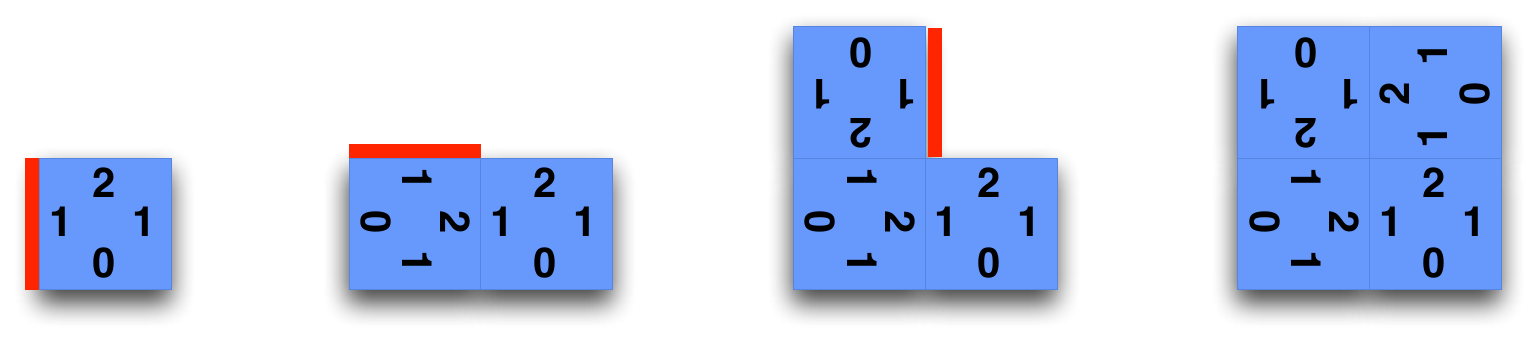}
      \caption{Example of steric occlusion for a cluster growing with tile $\{1,2,1,0\}$ under asymmetric interactions. The face randomly chosen at each step for subsequent attachment is shown in red. Note that the final cluster has no external 2-face left. Under single-seed growth conditions this cluster can continue to grow indefinitely as new single tiles with 2-faces continue to be available. However, under multi-seed conditions this cluster can only grow by binding to other clusters with free 2-faces.}\label{steric}
\end{figure}

In the following, we describe our computational model for multi-seed self-assembly of the same two tile sets $\{1,0,0,0\}$-$\{1,1,1,1\}$ and $\{1,2,0,0\}$-$\{1,2,1,0\}$. Firstly, note that clusters growing with tile set $\{1,2,0,0\}$-$\{1,2,1,0\}$ display at most a single free 2-face during growth. Once this 2-face is occupied, only 1-faces will be left on the perimeter. Such steric obstruction (see example in Figure \ref{steric}) is likely to happen in small cluster sizes ($N \sim 10$) for any $f$ value. In the single-seed case this is irrelevant, as new tiles are added one-by-one to a single cluster, and every new tile has a 2-face free for attachment. In the multi-seed case there will sooner or later be no single tiles left, and two clusters can only interact with each other if at least one of them has a free 2-face. This difference, which is a direct result of the asymmetric nature of the interaction, is the main reason why non-deterministic self-assembly of tiles with asymmetric interactions can lead to self-limiting growth. By adjusting the relative proportions of the two tiles $\{1,2,1,0\}$ and $\{1,2,0,0\}$ we can control the ratio of 0- and 1-faces on the perimeter of the clusters, which in turn allows us to regulate their average final size. Experimental confirmation (see below) of the self-limiting growth behaviour for tile set $\{1,2,0,0\}$-$\{1,2,1,0\}$ with asymmetric interactions, and the contrasting unbounded growth behaviour for tile set $\{1,0,0,0\}$-$\{1,1,1,1\}$ with symmetric interactions, shows that the computational model for the multi-seed initial condition is realistic. The details of the computational model for the multi-seed initial condition are set out below. The model differs for the two tile sets, and we describe the model for tile set $\{1,2,0,0\}$-$\{1,2,1,0\}$ first.

(a) Simulations start with an array of $n$ tiles, which can be thought of as clusters of size $1$. The variable $n$ can also be thought of as the total mass in the system.

(b) We randomly select two clusters, where the probability of selecting cluster $i$ is $p_i = n_i/n$, where $n_i$ is the size of cluster $i$. To account for different possible models of cluster selection we investigated more general probabilities $p_i \propto n_i^z$ with $0 \le z \le 3$, and found them to exhibit the same qualitative dependence on $f$ as the $p_i \propto n_i$ (i.e. $z=1$) case.

(c) There are three possibilities: (i) Neither cluster has a free 2-face, in which case step (b) is repeated, as the two clusters will not bind. (ii) One cluster has a free 2-face and the other does not, in which case the clusters bind, and the resulting cluster has no free 2-faces left. We then return to step (b). (iii) Both clusters have free 2-faces, in which case the clusters also bind. We then proceed to step (d) to decide whether the resulting cluster has a free 2-face.

\begin{figure}[h]
\includegraphics[width=0.5\textwidth]{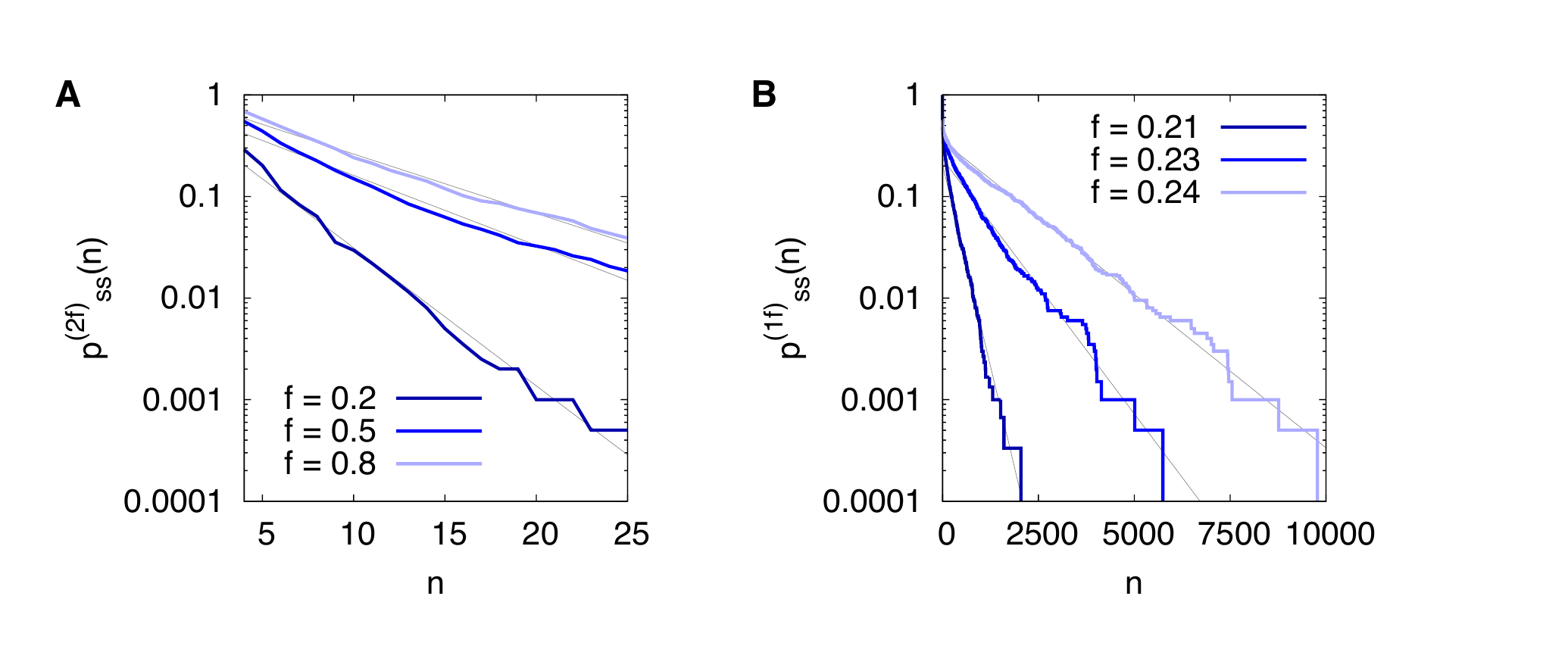}
\caption{A) The probability that a self-assembled cluster, grown using tile set $\{1,2,0,0\}$-$\{1,2,1,0\}$ under asymmetric interactions from the single-seed initial condition, has a 2-face on the perimeter, for $f = 0.2$ (dark blue), $0.5$ (medium blue), and $0.8$ (light blue). This probability distribution can be modelled very well as an exponential distribution (grey lines show best fit). B) The probability that a self-assembly cluster, grown using tile set $\{1,0,0,0\}$-$\{1,1,1,1\}$ under symmetric interactions from the single-seed initial condition, has a 1-face on the perimeter, for $f = 0.21$ (dark blue), $0.23$ (medium blue), and $0.24$ (light blue). This distribution can also be modelled very well by an exponential function (grey lines show best fit) for $f \le f_c = 0.25$, but for values of $f$ larger than $f_c$ this approximation breaks down, and a single giant cluster emerges.}\label{p2fp1f} 
\end{figure}

(d) In the single-seed growth model, the probability $p$ for a cluster to display a 2-face on its perimeter can be approximated by an exponential decay as a function of its size $n_i$, so that $p^{(2f)}_{ss} = e^{-cn_i}$ for which the decay constant $c$ depends on $f$, the relative density of the two tiles (see Figure \ref{p2fp1f}A). We approximate the merger of two clusters $i$ and $j$, with sizes $n_i$ and $n_j \ge n_i$, by a single-seed growth process that starts with the larger cluster $j$ and continues until that cluster has reached size $n_i+n_j$. The probability $p^{(2f)}_{ms}(n_i,n_j)$ for the newly formed cluster to still have a free 2-face is given by:
\begin{equation}\label{P2S}
p^{(2f)}_{ms}(n_i,n_j) = {e^{-c(n_i+n_j)}\over e^{-cn_j}}= e^{-cn_i}
\end{equation}
This is a conditional probability based on the single-seed growth approximation. The denominator corresponds to the fact that we started with a cluster of size $n_j$. If, after this step, there are then still clusters with free 2-faces, we return to step (b).

%
%

Upon termination of the growth process the clusters will exhibit a unimodal size distribution. To choose a characteristic cluster size we pick a cluster size $n_e$ such that all clusters smaller than $n_e$ have a total mass of $N/e$, or $1/e$ times the total mass. More formally, if the number of tiles in solution is $N$, and $s(n)$ is the number of clusters of size $n$, then:
\[
{N \over e} = \sum\limits_{n=1}^{n_e} n \, s(n)
\]
such that the $1/e$ percentile of the distributions corresponds to $n_e$. In the case of infinite growth the largest cluster can be larger than $N-N/e$ so that the size of the largest cluster becomes the characteristic size. We choose this procedure for calculating a characteristic cluster size from the distribution because it mirrors the way in which the characteristic cluster size is determined in the DNA tile experiments (see below), for which our model aims to offer a qualitative comparison.

For the tile set $\{1,0,0,0\}$-$\{1,1,1,1\}$, the same procedure as above is adopted with the difference that in (b-d) any two clusters with 1-faces can bind to each other. Hence, there is no extinction of 2-faces and we instead need to predict the extinction of 1-faces. To do so we consider the following model, similar to the above one:

(a) As above, simulations start with an array of $n$ tiles, which can be thought of as clusters of size $1$. The variable $n$ can also be thought of as the total mass in the system.

(b) As above, we randomly select two clusters, where the probability of selecting cluster $i$ is $p_i = n_i/n$, where $n_i$ is the size of cluster $i$. 

(c) If both clusters have at least one 1-face, continue to step (d). If not, return to step (b).

(d) Apply a similar method as above to determine the extinction of 1-faces. The probability $p^{(1f)}_{ms}(n_i,n_j)$ that the merged cluster still has a 1-face is derived by modelling the binding of the two clusters as the single-seed growth of the larger cluster, of size $n_j \ge n_i$, through the addition of $n_i$ single tiles, to a total size of $n_i+n_j$. The probability $p^{(1f)}_{ss}(n_i)$ that at least a single 1-face remains on a cluster of size $n_i$ grown under the single-seed initial condition can be obtained directly from single-seed growth simulations \cite{ARXIV2015}. Like $p^{(2f)}_{ss}(n_i)$, this probability distribution is also exponential, but only up to the critical value $f = f_c = 0.25$ (see Figure \ref{p2fp1f}B). Above this value of $f$, a single giant cluster emerges. To take into account that we started from the larger cluster of size $n_j$ we need to divide $p^{(1f)}_{ss}(n_i+n_j)$ by $p^{(1f)}_{ss}(n_j)$, to arrive at the probability distribution $p^{(1f)}_{ms}(n_i,n_j)$ needed for the decision whether the merged cluster has at least a single 1-face:
\begin{equation}\label{P1S}
\small
p^{(1f)}_{ms}(n_i,n_j) = {p^{(1f)}_{ss}(n_i+n_j) \over p^{(1f)}_{ss}(n_j)}
\end{equation}
The result of the two variants of the computational model are shown in Figure \ref{mainfig}A. For the tile set $\{1,2,0,0\}$-$\{1,2,1,0\}$ with asymmetric interactions (see Figure \ref{mainfig}A left) the final characteristic cluster size is relatively small and depends only weakly on $f$, illustrating the self-limiting growth behaviour of this tile set. For the tile set $\{1,0,0,0\}$-$\{1,1,1,1\}$ with symmetric interactions (see Figure \ref{mainfig}A right) we observe a transition from bound to unbound growth as a function of the relative tile concentration $f$. Note that these models have been tested across a wide range of ensemble sizes ($10^3$ to $10^7$ particles), and are entirely robust against this parameter. The simulation results shown in Figure \ref{mainfig}A have an ensemble size of $10^5$.

\section{Experimental results}

\begin{figure}[h]
\includegraphics[width=0.5\textwidth]{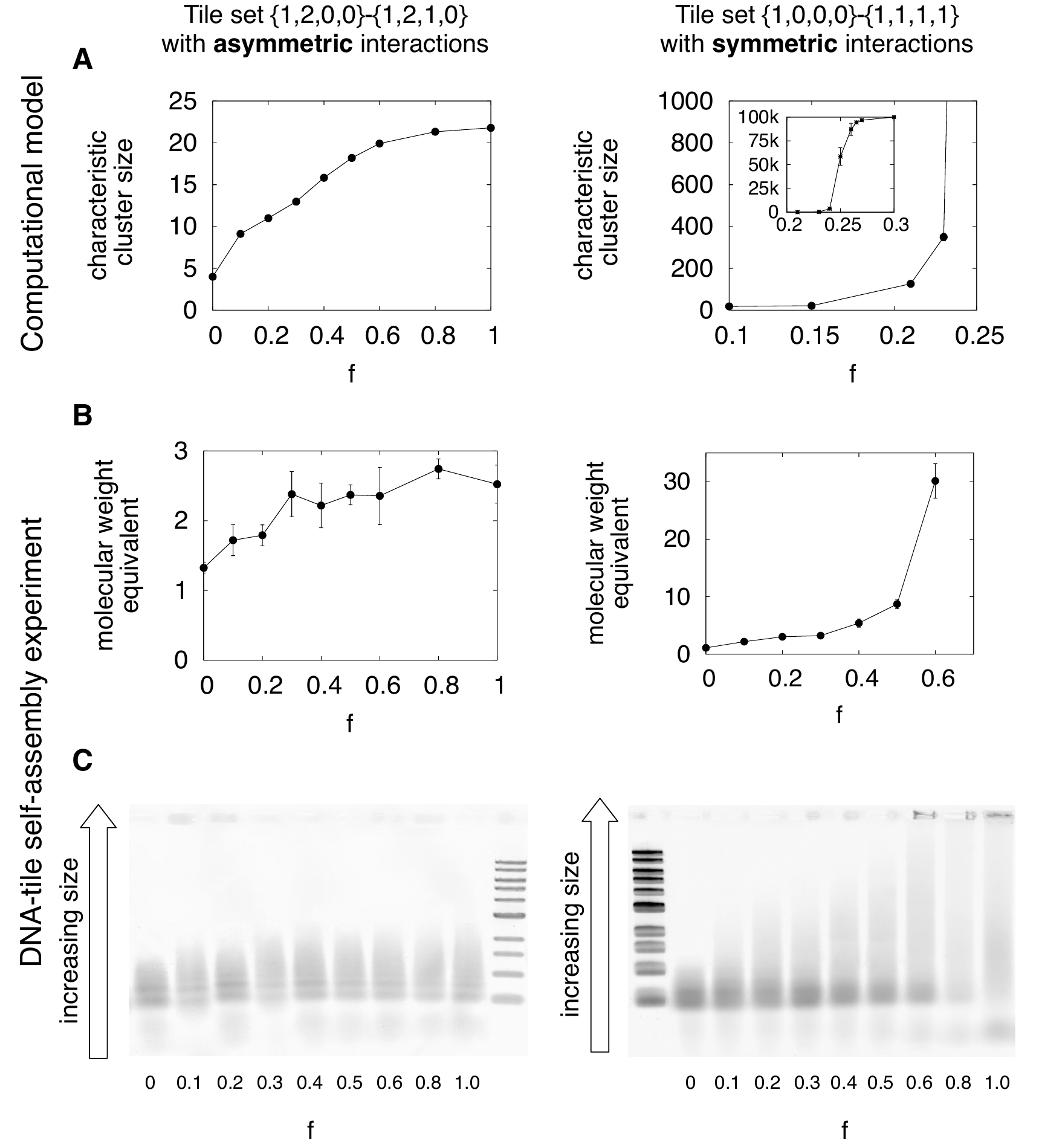}
\caption{Computational and experimental results for the two tile sets $\{1,2,0,0\}$-$\{1,2,1,0\}$ (left) and $\{1,0,0,0\}$-$\{1,1,1,1\}$ (right). A) Results of the computational multi-seed model, predicting self-limiting cluster growth for tile set $\{1,2,0,0\}$-$\{1,2,1,0\}$ with asymmetric interactions, and infinite growth of a single cluster for tile set $\{1,0,0,0\}$-$\{1,1,1,1\}$ with symmetric interactions. In the case of the asymmetric tile set the final cluster size is tunable by adjusting the relative concentrations of the two tiles. B) Results of the DNA tile self-assembly experiment, which match the computational prediction qualitatively. The asymmetric tile set shows self-limiting growth, while the symmetric tile set exhibits sharply increasing cluster sizes as $f$ increases. C) The agarose from which the experimental results (B) are derived. The values correspond to the intensity values at 5\% of the peak intensity, moving upwards through the gel.}\label{mainfig} 
\end{figure}

We build a physical realisation of the growth algorithm using self-assembling DNA-based tiles. While the specific base-pairing makes DNA ideally suited for duplication and storage of genetic information in biology, it has also been exploited to use DNA as a versatile and readily programmable material for efficient bottom-up fabrication at the nanoscale \cite{Seeman1983}. Applications of DNA origami \cite{Rothemund2006} and other types of DNA assemblies \cite{Wei2012d} include biosensors \cite{Bell2012,Hernandez-Ainsa2013}, nanoscopic rulers \cite{Steinhauer2009}, drug carriers \cite{Douglas2012b} or synthetic membrane channels \cite{Langecker2012a, Gopfrich2015}. Furthermore, addressable DNA lattices have been used as templates to design protein arrays \cite{Wei2012d, Park2006}. A four-armed junction with two parallel DNA duplexes per arm was the basic unit of these lattices. This design is ideally suited to implement our growth algorithm. \\
We equipped the basic four-armed DNA tile (see Figure \ref{Design} in the Appendix) with binding sites by extending both duplexes with 15 base long single-stranded overhangs on one (for the $\{1,0,0,0\}$ tile), two (for the $\{1,2,0,0\}$ tile), three (for the $\{1,2,1,0\}$ tile) or all four arms (for the $\{1,0,0,0\}$ tile). In order to ensure chirality for the asymmetric interactions, we used two different sequences on the two parallel helices. After assembly and characterisation of the DNA tiles (see Appendix), we mixed them at different ratios to cover a range of $f$-values from $0\leq f \leq 1$. To avoid DLA-like phenomena, we kept the concentration of tiles high (500 nM) and turned the interactions between the tiles on by adding sequence-complementary DNA linkers (see Appendix). \\
In order to assess cluster growth and to compare it to the algorithm, agarose gel electrophoresis was carried out. In an electric field, the negatively charged DNA migrates through a porous gel, whereby smaller clusters move faster than larger ones. The cluster size distribution is thus directly reflected by the intensity distribution along the gel visualised by an intercalating dye and UV-transillumiation (see Figure \ref{mainfig}C). There are apparent differences in the gels of the two tile sets: The smears extend much further down for the $\{1,0,0,0\}$-$\{1,1,1,1\}$ tile set, indicating the presence of larger clusters. For $f>0.6$, we observe aggregation in the gel loading pocket. This points towards the fact that clusters are larger than 100 nm, which is the typical pore size in a $1\%$ agarose gel \cite{Holmes1990}. These observations are supported by the dynamic light scattering traces showing mean hydrodynamic diameters above 250 nm for the $\{1,0,0,0\}$ tile at $f=1$ (see figure \ref{DLS} in the Appendix). For a more quantitative analysis of the gels, we determine the migration distance at which the intensity has dropped to $5 \%$ of its maximum and convert the value into a molecular weight equivalent by comparing it to the migration distance of DNA fragments of known length in the ladder.\\
In gel electrophoresis, the migration speed depends on charge and molecular weight, but also on the shape and the effective hydrodynamic radius of a DNA structure. Since a double strand of DNA is inherently different from our branched, more globular tile structures, it is unfortunately not possible to convert molecular weight equivalents to the number of tiles in a cluster \cite{Pluen1999}.


As shown in Figure \ref{mainfig}B, the obtained relation between $f$ and cluster size shows apparent differences for the two sets of tiles. While the cluster growth saturates at roughly $f=0.3$ for the $\{1,2,0,0\}$-$\{1,2,1,0\}$ tile set, the tile set with symmetric interactions continues to grow until it is too large to enter the gel. There is a 1.9 fold increase in the molecular weight equivalent from $f=0$ to $f=1$ for the tile set with asymmetric interactions, whereas the increase is 27-fold for the symmetric one from $f=0$ to $f=0.6$.
It should be noted, however, that a branched DNA structure migrates slower than the linear DNA fragments. Therefore, the molecular weight equivalent cannot directly be used to calculate the number of tiles within a cluster.
Our experimental data with DNA-based tiles thus shows a compelling qualitative agreement with the computational model. Despite experimental limitations and imperfections we can reproduce the two completely different growth phenomena in the two very similar sets of tiles.

\section{Discussion}

A multi-seed self assembly computational model of tiles $\{1,2,0,0\}$-$\{1,2,1,0\}$ under asymmetric interactions has produced quantitative calculations predicting a completely different behaviour compared to single-seed assembly. Multi-seeded assembly does not produce infinite clusters, whereas single-seed growth produces them beyond $f=0.53$. This phenomenon can be explained by the nature of the interactions between tiles, i.e. an asymmetric 1-2 coloured faces interaction with a tile set where tiles display at most a single 2-free face on each of them. The essential features of this model appear to be independent of the size of the system and the rate at which clusters stop growing. They are, however, strongly dependent on the rate at which clusters run out of 2-faces during assembly.

By contrast, multi-seed growth with tiles $\{1,0,0,0\}$-$\{1,1,1,1\}$ is qualitatively similar to the behaviour of the same tile set under the single-seed growth condition. This highlights the important role that interface asymmetry plays in the self-limiting character of the growth of tile set $\{1,2,0,0\}$-$\{1,2,1,0\}$ from multiple seeds.

Multi-seed conditions are more realistic, both biologically and experimentally. However, single-seed conditions could in principle be imposed in a DNA tile experiment by carefully regulating the availability of new tiles, so that new tiles bind one at a time. Aside from vastly increasing the time scale of the experiment, this procedure would likely also lead to DLA-like dynamics, which would have to by mitigated by additional modifications of the experimental design. Similarly, and more feasibly, the size of aggregates could be further tuned by making building blocks available in swathes, which would likely lead to a multimodal cluster size distribution. 

The DNA tiles are much less rigid in shape than the two-dimensional square tiles of the computational model, which means that the growth of DNA tiles is not strictly confined to two dimensions. This is likely to account for the differences in the $f$ values at which the cluster size diverges for the tile set $\{1,0,0,0\}$-$\{1,1,1,1\}$. The qualitative behaviours of both tile sets however are unaffected by this difference in tile shape, underlining the robustness of the computational model, and of its results. 
\newline

\section{Conclusions}

We show here, both through a computational model and through DNA self-assembly experiments, that a mixture of two self-assembly building blocks with asymmetric interfaces can lead to tunable self-limiting cluster growth. As protein interfaces are often asymmetric, this model informs the study of protein aggregates, which can result from misfolded proteins that would otherwise build a particular structure, such as a protein complex, deterministically.

In addition, these results lead to applications in the field of DNA-based bioengineering, or might suggest new uses for inorganic nanoparticles with asymmetric interfaces. Further formal investigation of other tile sets under multi-seed conditions, building upon the comprehensive study of single-seed growth behaviours in \cite{ARXIV2015} with computational models and DNA self-assembly experiments, is also likely to prove fruitful in this context.

\section{Acknowledgments}

K.G. acknowledges funding from the Winton Programme for the Physics of Sustainability, Gates Cambridge and the Oppenheimer PhD studentship, T.K. and K.G. from the NanoDTC Cambridge EP/L015978/1. S.T. acknowledges support form the UK Engineering and Physical Sciences Research Council (EP/L504920/1). S.E.A. acknowledges support from The Royal Society.

\section{Appendix}

\section{Methods}

\subsection{Assembly of the DNA-based tile monomer}
\begin{figure}[h]
\begin{center}
\includegraphics[scale=1]{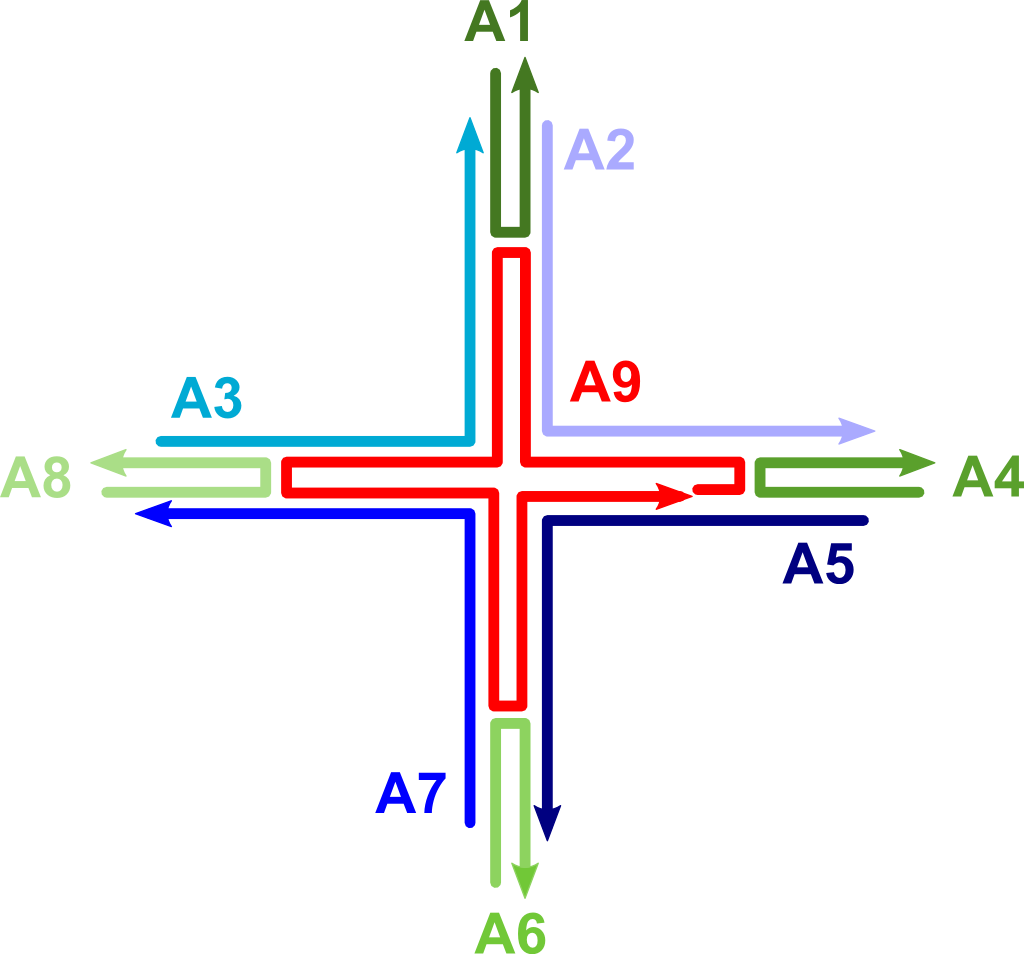}
\end{center}
\caption{\label{Design} Pathways of DNA strands for a tile monomer (design adapted from \cite{Park2006,Yan2003}). Arrows indicate 3' ends of the DNA strands.}
\end{figure}

Our DNA tiles are based on an earlier design by Yan et al.~\cite{Yan2003} and Park et al.~\cite{Park2006}. It is a four-armed junction with two parallel DNA duplexes per arm (see Figure \ref{Design}). Each arm is $7.5\unit{nm}$ long, resulting in a total width of $15\unit{nm}$. We equipped the reactive arms of our DNA tile monomer with 15 base long single-stranded DNA overhangs complementary to the sequence of a DNA linker (see Table 1), while the non-reactive arms carried non-sticky ends to prevent unspecific interaction with other tiles \cite{Yin2008a}. 
All nine DNA sequences of a monomer (Integrated DNA Technologies, Inc.) were mixed at equimolar ratios and diluted to a final concentration of $500\unit{nM}$ in $10\unit{mM}$ Tris-HCl, $1\unit{mM}$ EDTA, $50\unit{mM}$ $\mathrm{MgCl}_2$, pH 8.0. The mixture was annealed using a standard protocol (heating to $80\deg$ for 5 min, cooling down to $65\deg$ using a linear cooling ramp over 75 min, subsequently cooling to $25\deg$ within $16\unit{h}$ in a thermocycler (BioRad)). 

\onecolumngrid
\begin{longtable}{|l|l|}
\hline \textbf{Sequence Name} & \textbf{Sequence (5' to 3')} \\ 
\hline A2 & TGAT GATG CAAC CTGC CTGG CAAG ACTC CAGA GGAC \\ & TACT CATC CGT\\ 
\hline A3 & GTGG AATA GCGC CTGA TCGG AACG CCTA CGAT GGAC \\ & ACGC CGCT ACC\\
\hline A5 & TCCG ACTG AGCC CTGC TAGG ATCG ACTT CACT GGAC \\ & CGTT CTAC CGA\\
\hline A7 & ACCG GAGG CTTC CTGT ACGG CAGA ACTC CGTT GGAC \\ & GAAC AGTG AGT\\
\hline A9 & AGGC ACCA TCGT AGGT TTTC GTTC CGAT CACC \\ & AACG GAGT TTTT TCTG  CCGT ACAC CAGT GAAG \\ & TTTT TCGA TCCT AGCA CCTC TGGA GTTT TTCT TGCC\\
\hline A1-onearm & TTTT TGGT AGCG GCGT GTGG TTGC ATCA TCAT TTTT\\
\hline A4-onearm-reactive & CTTG GTCG ACTC AGGA CGGA TGAG TAGT GGGC TCAG \\ & TCGG AGTA CCTC GGGT ACCA\\
\hline A6-onearm & TTTT TTCG GTAG AACG GTGG AAGC CTCC GGTT TTTT\\
\hline A8-onearm & TTTT TACT CACT GTTC GTGG CGCT ATTC CACT TTTT\\
\hline A1-twoarm-reactive & ACTT ACTC AGGT TATG GTAG CGGC GTGT GGTT GCAT \\ & CATC AGAG CTCG AGTG TGTC\\
\hline A4-twoarm-reactive & CTTG GTCG ACTC AGGA CGGA TGAG TAGT GGGC TCAG \\ & TCGG AGTA CCTC GGGT ACCA\\
\hline A6-twoarm & TTTT TTCG GTAG AACG GTGG AAGC CTCC GGTT TTTT\\
\hline A8-twoarm & TTTT TACT CACT GTTC GTGG CGCT ATTC CACT TTTT\\
\hline A1-threearm-reactive & ACTT ACTC AGGT TATG GTAG CGGC GTGT GGTT GCAT \\ & CATC AGAG CTCG AGTG TGTC\\
\hline A4-threearm-reactive & CTTG GTCG ACTC AGGA CGGA TGAG TAGT GGGC TCAG \\ & TCGG AGTA CCTC GGGT ACCA\\
\hline A6-threearm & TTTT TTCG GTAG AACG GTGG AAGC CTCC GGTT TTTT\\
\hline A8-threearm-reactive & CTTG GTCG ACTC AGGA CTCA CTGT TCGT GGCG CTAT \\ & TCCA CGTA CCTC GGGT ACCA\\
\hline A1-fourarm-reactive & CTTG GTCG ACTC AGGG GTAG CGGC GTGT GGTT GCAT \\ & CATC AGTA CCTC GGGT ACCA\\
\hline A4-fourarm-reactive & CTTG GTCG ACTC AGGA CGGA TGAG TAGT GGGC TCAG \\ & TCGG AGTA CCTC GGGT ACCA\\
\hline A6-fourarm-reactive & CTTG GTCG ACTC AGGT CGGT AGAA CGGT GGAA GCCT \\ & CCGG TGTA CCTC GGGT ACCA\\
\hline A8-fourarm-reactive & CTTG GTCG ACTC AGGA CTCA CTGT TCGT GGCG CTAT \\ & TCCA CGTA CCTC GGGT ACCA\\
\hline Linker-onefour & CCTG AGTC GACC AAGT GGTA CCCG AGGT AC\\
\hline Linker-twothree-1 & CCTG AGTC GACC AAGG ACAC ACTC GAGC TC\\
\hline Linker-twothree-2 & CATG GAGC CCAT GGTT GAAT GAGT CCAA TA\\

\hline
\caption{DNA sequences used for the assembly of the DNA tiles.}
\end{longtable}
\twocolumngrid

In order to ensure chirality, we use two different linkers for the tile set with assymmetric interactions. A single linker sequence is sufficient for the other tile set, since chirality does not matter here.

\subsection{Characterisation via polyacrylamide gel electrophoresis, UV-melting and dynamic light scattering}\label{carcar}
\begin{figure}[H]
\begin{center}
\includegraphics[scale=1]{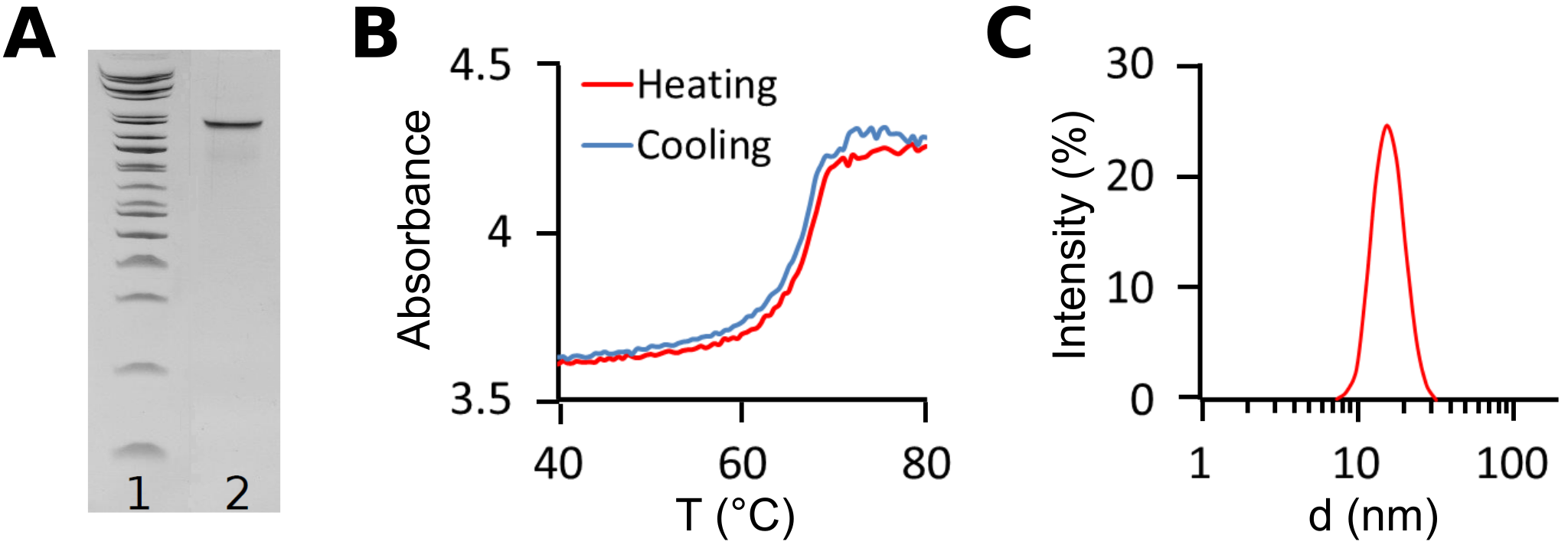}
\end{center}
\caption{\label{Monomer}A) $10\%$ Polyacrylamide gel electroporesis of a 2-log DNA ladder (Lane 1) and a tile monomer composed of 9 single-strands of DNA (Lane 2). B) UV-melting profile of the DNA tile monomer. C) Dynamic light scattering trace of the DNA tile monomer. }
\end{figure}

The DNA tile monomers were characterised using $10\%$ polyacrylamide gel electrophoresis in a solution containing $11\unit{mM}$ MgCl$_2$ buffered to approximately pH 8.3 with $45\unit{mM}$ Tris-borate, $1\unit{mM}$ EDTA and running conditions of $160\unit{V}$ and 90 min. Bands were stained with GelRed and visualised using UV-transilluminaton. The gel yielded a single sharp band pointing towards uniform structure size and shape (Figure \ref{Monomer}A).  \\
As an additional confirmation for the uniform assembly, UV-melting experiments were carried out. $500\unit{nM}$ of the DNA tiles were added to a quartz cuvette (Sigma Aldrich). The mixture was subjected to heating and cooling cycles ($25\deg$ to $80\deg$) at a rate of $0.25\deg$/min in a Varian Cary 300  UV/Vis spectrophotometer while monitoring the absorption at $260\unit{nm}$. As expected, a sharp melting transition is observed with a melting point at $68\deg$ (see Figure \ref{Monomer}B). \\
Dynamic light scattering (DLS) was carried out on a Zetasizer Nano S (Malvern) to probe the structure size. The measured hydrodynamic diameter of $16.9\pm 0.03\unit{nm}$ is in good agreement with the theoretical size of the DNA tile monomers ($15\unit{nm}$ across, see Figure \ref{Monomer}C).

\subsection{Linking the monomers for cluster growth}
After assembly, the monomers were incubated with linkers (30 base long DNA sequences complementary to the respective single-stranded DNA overhangs of the monomer) for 24 hours at room temperature. The concentration of the linkers was selected to saturate statistically half of the binding sites to achieve efficient linking. 
Agarose gel electrophoresis and dynamic light scattering were used to test whether the linking was successful.

\subsection{Dynamic light scattering of tile clusters}
\begin{figure}
\begin{center}
\includegraphics[scale=0.8]{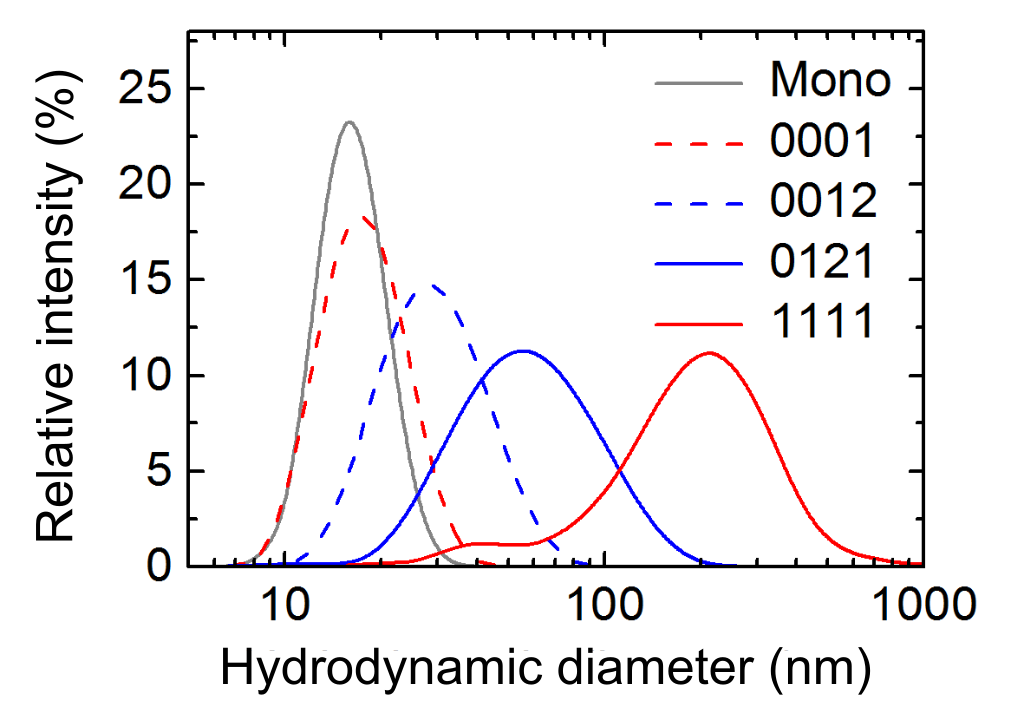}
\end{center}
\caption{\label{DLS} Dynamic light scattering traces for the DNA tile monomer (grey, solid), the $\{1,0,0,0\}$ tile (red, dashed), the $\{1,2,0,0\}$ tile (blue, dashed), the $\{1,2,1,0\}$ tile (blue, solid) and the $\{1,1,1,1\}$ tile (red, solid). }
\end{figure}

Since we expect the distribution of tile cluster size to be heterogeneous, we cannot draw quantitative conclusions  based on the dynamic light scattering experiments. However, the data presented in Figure \ref{DLS} shows that linking has taken place and that the sizes change for different tiles.
The monomer shows a narrow size distribution with a mean hydrodynamic diameter of $16.95 \pm 0.03\unit{nm}$. The $\{1,0,0,0\}$ tile, which is expected to form dimers, has a slightly larger hydrodynamic radius of $19.11 \pm 0.06\unit{nm}$. The $\{1,2,0,0\}$ tile which should form quadrumers is even larger with a mean diameter of $33.23 \pm 0.13\unit{nm}$. The $\{1,2,1,0\}$ tile reaches a mean diameter of $73.26 \pm 0.47\unit{nm}$, while the $\{1,1,1,1\}$ tile even reaches $250.99 \pm 3.91\unit{nm}$. While the mean hydrodynamic diameter of the $\{1,2,0,0\}-\{1,2,1,0\}$ tile set increases by a factor of 2.2 from $f = 0$ (only $\{1,2,0,0\}$) to $f = 1$ (only $\{1,2,1,0\}$), it increases by a factor of 13.1 for the tile set with symmetric interactions.  
The means and standard errors were determined from log-normal fits of the distributions. 

\subsection{Agarose gel electrophoresis of tile clusters}
Agarose gel electrophoresis was used to obtain the cluster size distributions. A $1\%$ gel was run for $3\unit{h}$ at $60\unit{V}$ in a solution containing $11\unit{mM}$ MgCl$_2$ buffered to approximately pH 8.3 with $45\unit{mM}$ Tris-borate, $1\unit{mM}$ EDTA. Bands were stained with GelRed and visualised using UV-transilluminaton. A 1 kilobasepair (kbp) DNA ladder (New England Biolabs) was used as a molecular weight reference.

\subsection{Data extraction from agarose gels}

\begin{figure}
\begin{center}
\includegraphics[scale=0.8]{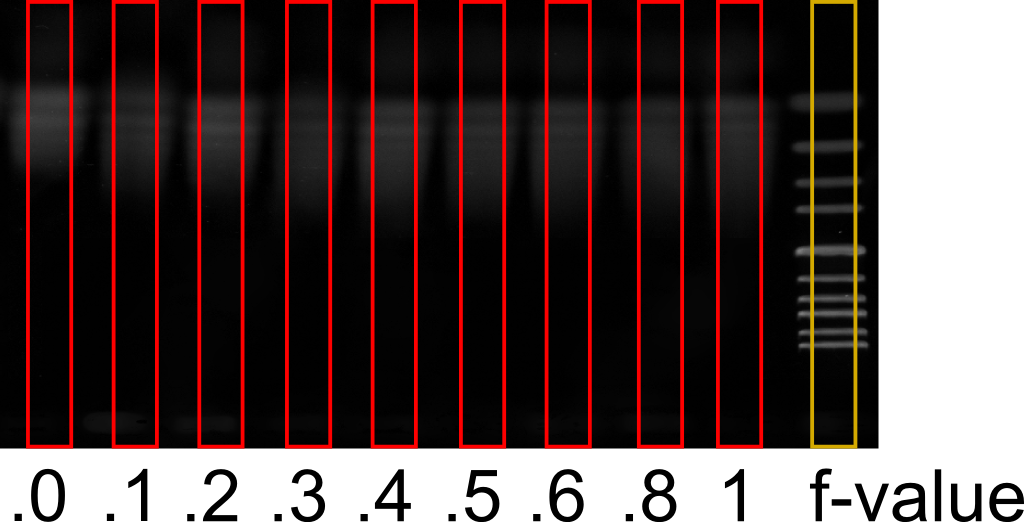}
\end{center}
\caption{\label{Interestregion} Exemplary agarose gel of the 0012/0121 tile set. Regions of interested for f-values from 0 to 1 are highlighted in red. The region of interested for the 1 kbp ladder is highlighted in yellow. }
\end{figure}

\begin{figure}
\begin{center}
\includegraphics[scale=0.3]{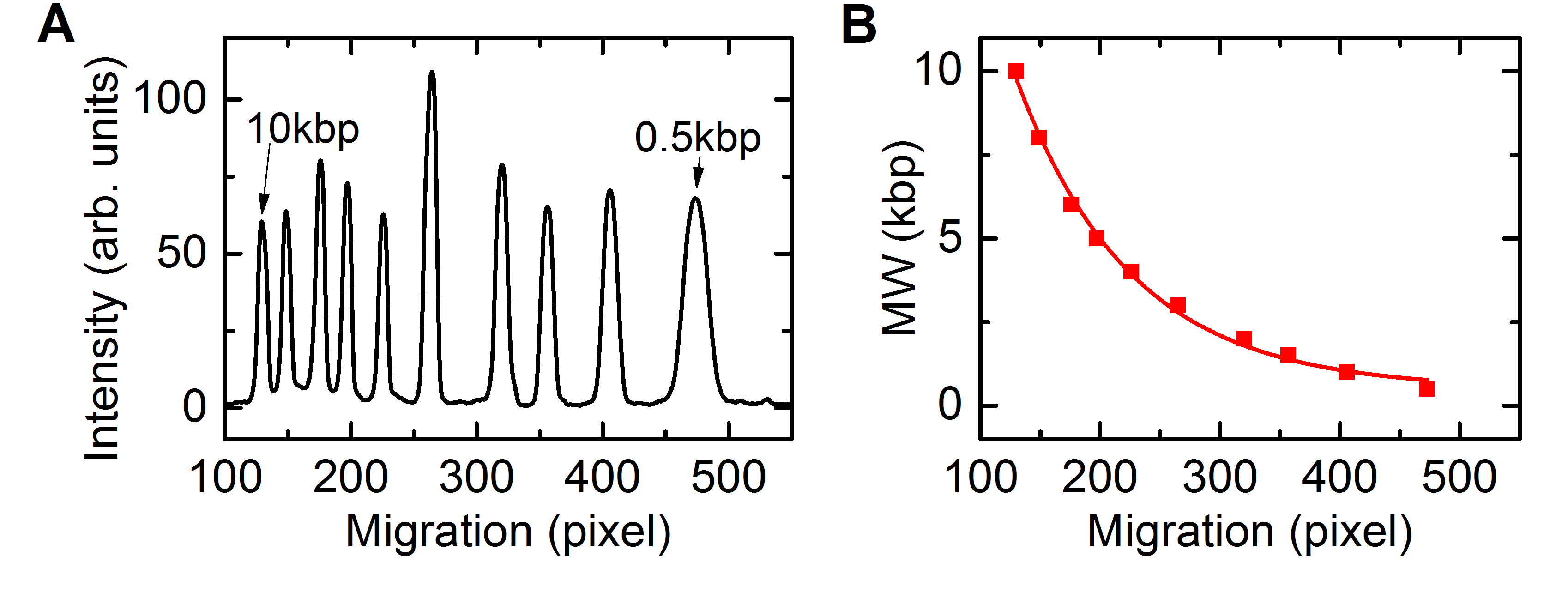}
\end{center}
\caption{\label{ladderfit} A) Intensity profile of the 1 kilobase pair (kbp) DNA ladder. Each peak corresponds to a DNA fragment of known length. B) Molecular weight (MW) as a function of migration distance as extracted from the intensity profile in A. The data has been fitted with an exponential decay (red line). }
\end{figure}

Data extraction from the agarose gel images was performed as follows:
Images were imported into \textit{ImageJ}. After background subtraction, a 40 pixel wide region of interest was defined for each lane in the gel as shown in the example image in Figure \ref{Interestregion}. Intensity profiles (averaged across the 40 pixel region) were generated for each lane. The peaks of the intensity profile of the 1 kbp DNA ladder (see Figure \ref{ladderfit}A) correspond to a double-stranded DNA segment of known length between 0.5 ad 10 kilobasepairs (kbp). An exponential decay was fitted to the peak values to obtain the relation between migration distance (in pixel) and molecular weight (in kbp) as shown in Figure \ref{ladderfit}B. Even though a branched cross structure is expected to migrate slower than an linear DNA double-strand with the same number of base pairs, we can use the ladder as a molecular weight equivalent. It is important to note that this quantity can not directly be used to determine the number of crosses within a cluster.

\begin{figure}
\begin{center}
\includegraphics[scale=1.1]{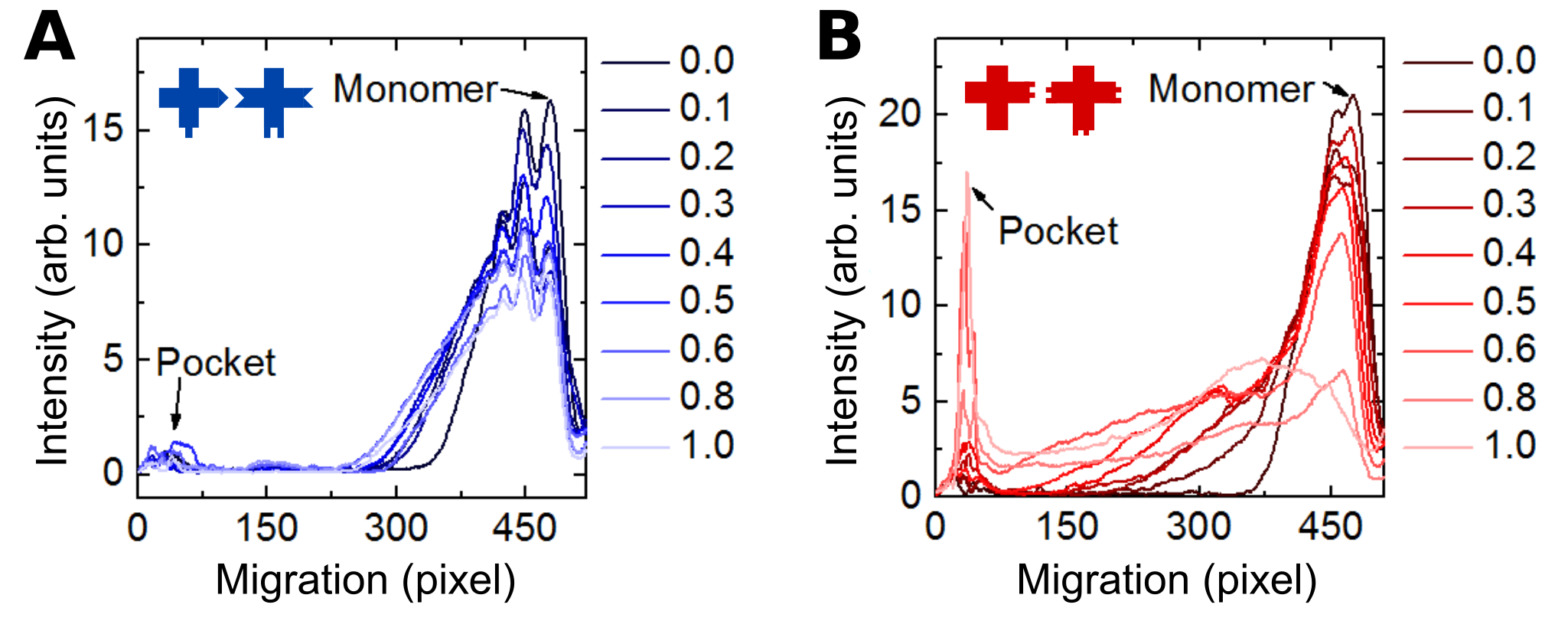}
\end{center}
\caption{\label{Intensities} Intensity distributions as a function of migration distance extracted from agarose gel electrophoresis images for the tile set with A) assymetric interactions and B) symmetric interactions. The legend shows the colour-coding for different f-values.}
\end{figure}

The intensity profiles of the tile clusters imported in \textit{Matlab}. A larger migration distance corresponds to small cluster size. We determine the migration distance at which the intensity has dropped to $5 \%$ of its maximum and convert the value into a molecular weight equivalent by comparing it to the DNA ladder and using the relation determined before. In Figure 4B in the main text we then plotted this molecular weight equivalent against the f-values we tested. The intensity profiles for the two tile sets and all tested f-values are shown in Figure \ref{Intensities}. Almost all of them show a peak at the migration distance corresponding to a single tile monomer. In order to avoid this peak and to ensure we take large clusters into account, we chose the very low $5\%$ threshold. \\
The fact that we observe monomers reflects the imperfections of the experimental system, such as misfolded structures or non-optimal conditions for linking. Since we are adding the single-stranded DNA linkers after assembly of the crosses, two crosses that have already bound linkers are less likely to attach to one another. Given these experimental imperfections the good qualitative agreement between theory and experiments is even more impressive. The theory proves to be robust.

\bibliographystyle{apsrev4-1}
\bibliography{PRreferences3}
\end{document}